\documentclass[aps,twocolumn,showpacs]{revtex4-1}
\usepackage{dcolumn}
\usepackage{bm}
\usepackage{amssymb}
\usepackage{amsfonts}
\usepackage{amsmath}
\usepackage{graphicx}

\begin{document}

\title{Mutagenesis and Background Neutron Radiation}
\author{Augusto Gonz\'alez}
\affiliation{Instituto de Cibern\'etica, Matem\'atica y F\'{\i}sica, La Habana}

\begin{abstract}
We suggest a possible correlation between the ionization events  caused by the background neutron radiation and the experimental data on mutations with damage in the DNA repair mechanism, coming from the  Long Term Evolution Experiment in E. Coli populations. 
\end{abstract}

\maketitle

\section{Introduction}

In microelectronics, single failure events sporadically occur which, in some areas, like plane and space navigation, could have catastrophic consequences. Preliminary estimations \cite{IBM} and more recent experiments \cite{exp} indicate a correlation between these events and the Background Neutron Radiation (BNR) \cite{RFN}. The mechanism of failure is the collision of a neutron from the BNR with an atomic nucleus in the chip, leading to a shower of electrons and ions that locally changes the conductivity and shortcuts the device.

In the present paper, we suggest the BNR as a cause of genetic ``fails'' in living cells, that is one of the possible origins of the so called spontaneous mutations. Cells exposed to the shower of electrons and ions, caused by the collision of a neutron and a proton of water, could be anihilated or experience a permanent damage, in particular, a damage in the DNA. The frequency of such events is similar to the rate of appearance of mutations with damage in the DNA repair mechanism \cite{Lenski}, as measured in the Long Term Evolution Experiment (LTEE), where E. Coli populations evolve under controlled conditions \cite{EELP}.

\section{The LTEE in E. Coli populations}
\label{EELP}

The LTEE is an experiment conduced by Prof. R. Lenski and his group at the Michigan State University \cite{EELP}. Each day, the bacteria undergo 6 - 7 generations of binary evolution. In a year, around 3400 generations occur. This means that, since the experiment started in 1988, it passed 60000 generations.

\begin{figure}[ht]
\begin{center}
\includegraphics[width=0.9\linewidth,angle=0]{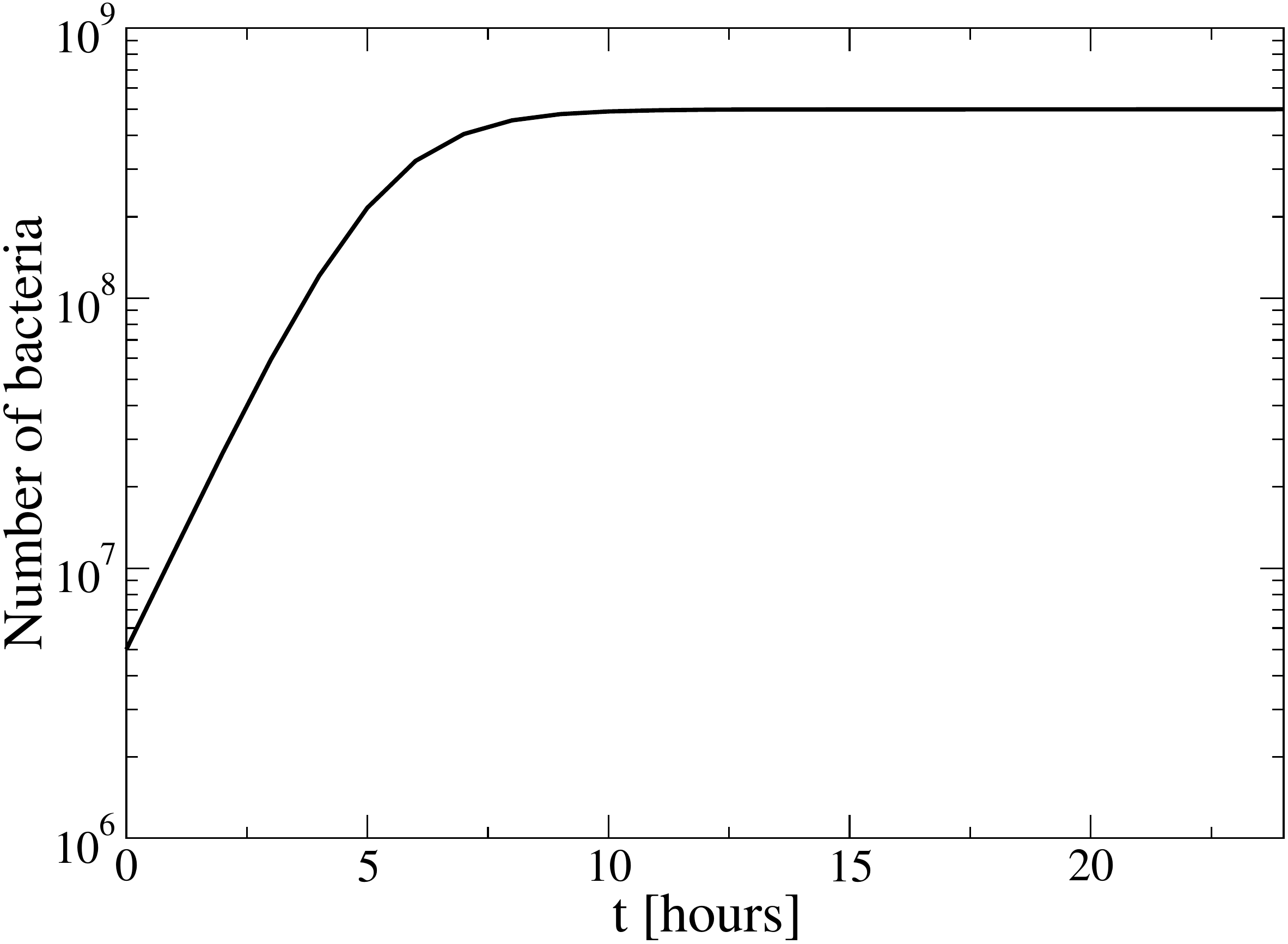}
\caption{Daily evolution of the number of bacteria in a culture in the LTEE.}
\label{fig1}
\end{center}
\end{figure}

In the experiment, 12 populations of bacteria, with a common ancestor, independently evolve. Every day, 0.1 ml of the bacterial culture is serially transferred to 9.9 ml of a glucose solution, and mantained under controlled temperature until the next day. The number of bacteria varies approximately as shown in Fig. \ref{fig1}. That is, grows according to the law $N_0~2^{t/t_0}$ in the first 8h, until the glucose is depleted, and then reach a stationary state. In the last 16h there is no appreciable mortality. The dependence $2^{t/t_0}$ is due to the way of reproduction, by cellular division.

The experiment shows a set of very interesting results \cite{Lenski}. We shall stress only two of them. First, in a given population, the total number of single point mutations in the DNA, after 20000 generations of evolution, is estimated as $3\times 10^8$. That is, the rate of point mutuations is:

\begin{equation}
f_{SPM}\approx 1~s^{-1}.
\label{fspm}
\end{equation}

On the other hand, in 2 of the 12 cultures, after 2500 - 3000 generations, mutations with a damaged DNA  repair and edit mechanism appeared and became numerically dominant. A third line evolved the mutator phenotype after 8500 generations, and a fourth after 15000 generations. According to Prof. Lenski, the mechanism through which the mutator becomes numerically dominant is roughly the following. Once the mutation appears, the rate of spontaneous mutations increases 100 times, as compared with cells in which the DNA repair mechanism is not damaged. Thus, the mutator has a higher chance to generate the next winner and become dominant in a relatively short time scale, around 250 generations.

A second aspect, stressed by Prof. Lenski, is that mutations in which the DNA repair mechanism is damaged are ``deleterious'', in the sense that a segment of the DNA is removed.

\section{Ionization events caused by the BNR}

With regard to the BNR, we may assume that the cells live in pure water. Indeed, water is the main component of the solution, and the pH should be close to 7 in order to preserve life \cite{pH}. In these conditions, the important processes are  the collisions between neutrons, from the BNR, and the Hydrogen nuclei (protons) of water. The ejected proton gives rise to a shower of ions and electrons that is extended approximately 0.1 mm along the proton trajectory.

\begin{figure}[ht]
\begin{center}
\includegraphics[width=0.9\linewidth,angle=0]{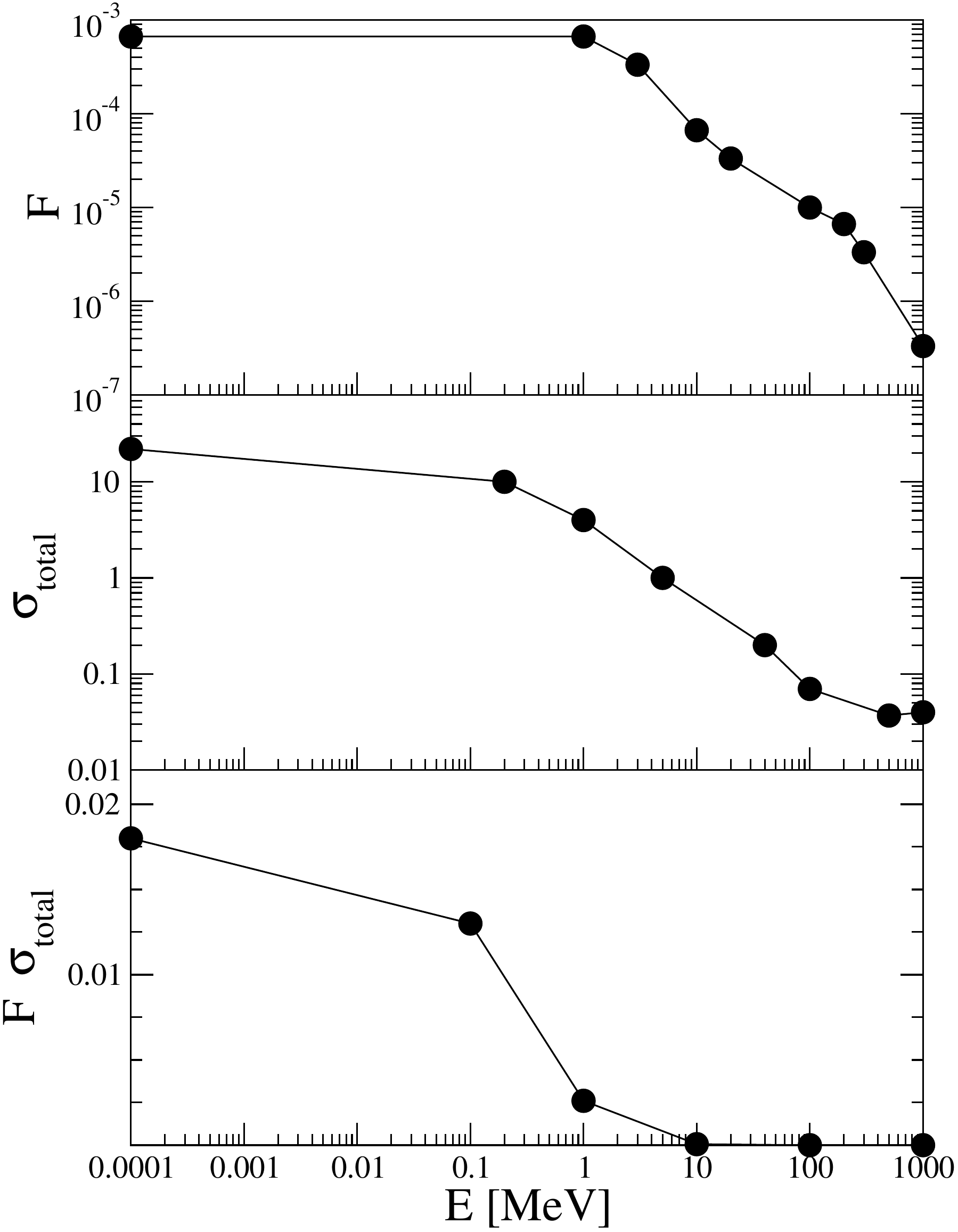}
\caption{Neutron flux per unit of energy in the BNR, $F$, total cross section of the n+p collision, $\sigma_{total}$, and the product $F~\sigma_{total}$, as functions of the energy of the incident neutron.}
\label{fig2}
\end{center}
\end{figure}

In Fig. \ref{fig2}, we show the flux per unit energy of neutrons in the BNR \cite{RFN}, $F$, in units of Neutrons/(MeV s cm$^2$); the  total cross section for neutron-proton dispersion \cite{sigma}, $\sigma_{total}$, in units of 10$^{-24}$ cm$^2$; and the product $F~\sigma_{total}$, in units of 10$^{-24}$ s$^{-1}$ MeV$^{-1}$, as functions of the energy of incident neutrons. This last magnitude is proportional to the probability that a neutron with a given energy collides with a proton in water. It can be noticed that only neutrons with energy lower than a few MeV have a significant effect. 

From these data, we may estimate the probability of neutron-proton  collisions:

\begin{eqnarray}
Prob_{n+p}&=& N_p \int_0^{1000~MeV} {\rm d}E~F~\sigma_{total}\nonumber\\
          &=&8\times 10^{-3} s^{-1},
\end{eqnarray}

\noindent
where $N_p=6.6\times 10^{23}$ is the number of protons in 10 ml of water.

\begin{figure}[ht]
\begin{center}
\includegraphics[width=0.9\linewidth,angle=0]{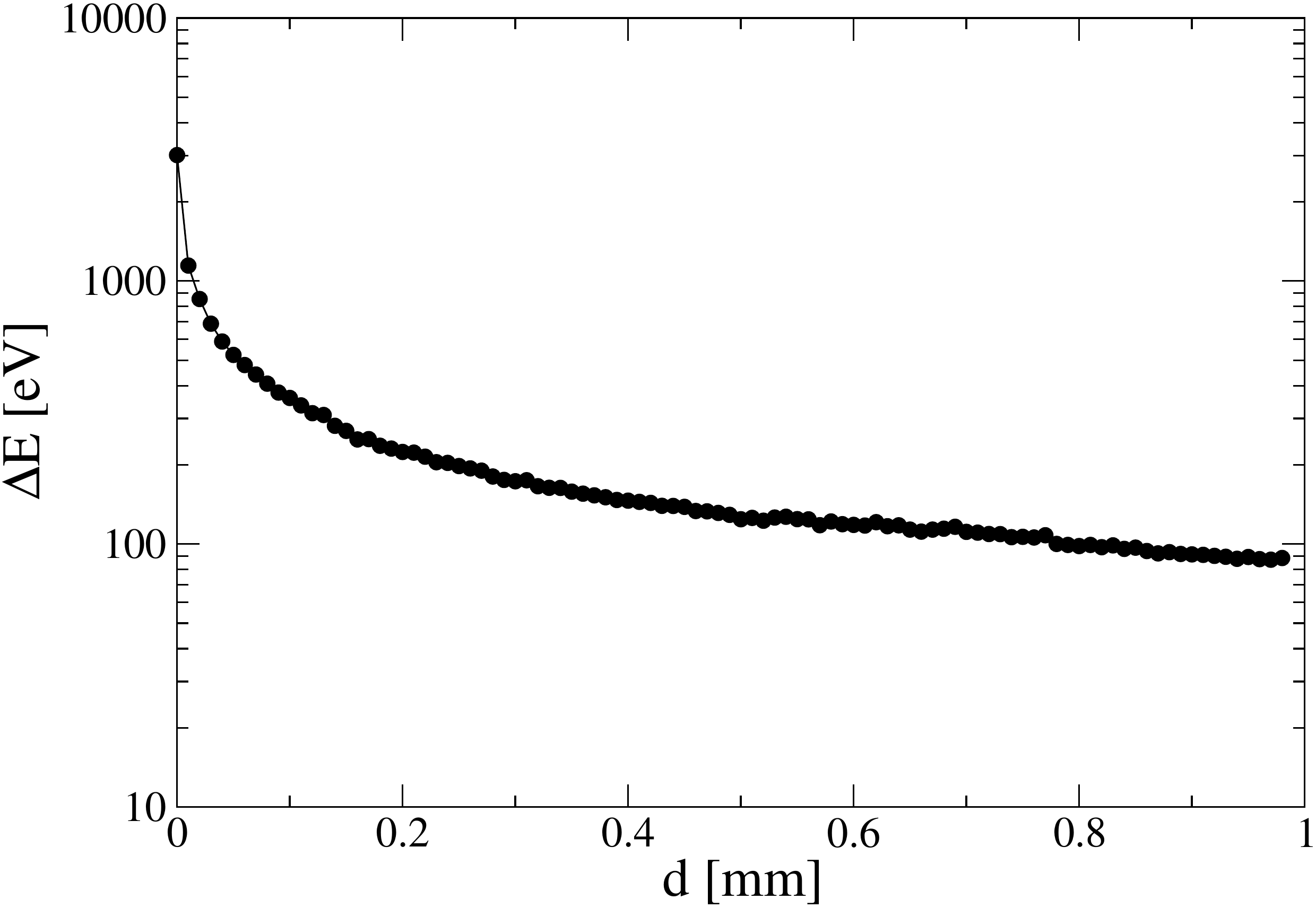}
\caption{Energy transfer from the proton to the medium, in 100 nm steps, along the proton trajectory.}
\label{fig3}
\end{center}
\end{figure}

In order to compute the energy transfer from the proton to water, we use the data in Fig. \ref{fig2} and the so-called stopping power of protons in water, tabulated in Ref. \onlinecite{sp}. By using a Monte Carlo algorithm, we arrive to the results shown in Fig. \ref{fig3}. According to Ref. \onlinecite{sp}, energy losses are mainly due to the interaction of the proton with the electrons in the medium, leading to the ionization of water molecules. The basic process of ionization is: ${\rm H_2O\to e+H_2O^+}$, which requires an energy of 12.62 eV \cite{Ip}. The ejected electron and ${\rm H_2O^+}$ could lead to secondary ionization processes. Dividing the $y$ axis of Fig. \ref{fig3} by 10, we  obtain a rough estimate of the number of ions produced in each 100 nm step of the proton motion, that is around 300 ions at distances close to the n-p collision point, and 30 ions when distance is of the order of 0.1 mm.

\section{BNR effects on the LTEE}

As mentioned above, n+p dispersion events in the glucose solution take place every 125 s. We already know that the shower of ions and electrons, created by the proton, is more intense in the first 0.1 mm along the proton trajectory. The bacteria touched by this ion shower could be destroyed or experience a permanent damage, especially in their DNA, which can be later inherited by the descendants. We shall stress that, for the DNA changes to be transmitted, the ionization event should take place in the first 8h of daily evolution, according to Fig. \ref{fig1}. Otherwise, there is practically no cellular division in the day it occurred, and the probability to pass to the next day is only 1/100.

The mean number of bacteria in the first 8h is:

\begin{equation}
\bar N=\frac{N_0}{8}\int_0^8 {\rm d}t~2^{t/t_0}=21.5~N_0,
\end{equation}

\noindent
where $N_0=5\times 10^6$ bacteria, and $2^{8h/t_0}=100$. Each bacterium occupies a mean volume of around 10 cm$^3/(21.5~N_0)$, that is, a cube with sides 45 $\mu$m long. In the first 0.1 mm=100 $\mu$m of the ion shower, only 2 such cubes could be allocated. The probability that the shower touches a bacterium is, thus:

\begin{equation}
2~\frac{\rm Shower~Volume}{\rm Cube~Volume}=2~\frac{l^2\times 45 ~\mu m}{(45 ~\mu m)^3},
\label{vol}
\end{equation} 

\noindent
where $l$ is the lateral dimension of the ion shower. $l$ could be estimated from the Debye screening length of pure water:

\begin{equation}
\lambda_D=\left(\frac{k_B T\epsilon\epsilon_0}{2 n q^2}\right)^{1/2},
\end{equation}

\noindent
where $k_b$ denotes the Boltzman constant, $\epsilon\approx 80$ is the relative dielectric constant of water \cite{epsilon}, $q=1$ is the charge of the ions H$^+$ y OH$^-$ in water, and $n$ their concentration:

\begin{eqnarray}
n&=&10^{-7}\times (3\times 10^{22}~{\rm molecules/cm^3})\nonumber\\
&=&3\times 10^{15} ~{\rm ions/cm^3}.
\label{n}
\end{eqnarray}

\noindent
Taking all these numbers together, we get $\lambda_D=500~nm=0.5~\mu m$. And putting $l=\lambda_D$ in Eq. (\ref{vol}), we get a probability of $2.5\times 10^{-4}$. Notice that $l$ is a magnitude of the same order of the E. Coli dimensions, thus the ion shower may cause strong effects on a bacterium.

We may compute the rate in which bacteria from a single population are touched by the BNR ionization events:

\begin{eqnarray}
f_{BNR}&\approx& (2.5\times 10^{-4})\times (8\times 10^{-3}~s^{-1})\nonumber\\
&\approx& 2\times 10^{-6} s^{-1}.
\end{eqnarray}

\noindent
This number is very small, as compared with $f_{SPM}$, Eq. (\ref{fspm}). However, it is consistent with the frequency of deleterious mutations, with damage in the DNA repair mechanism, mentioned in section \ref{EELP}. Indeed, in $\Delta t\sim 2400$ generations $\sim 1$ year $\sim 3\times 10^7$ s, the BNR had a direct incidence on $f_{BNR}\times \Delta t\sim 60$
bacteria. Some of them could have experienced damages in the DNA repair mechanism. The 100 times increase in the mutation rate could have given this subpopulation, after 100 - 600 generations, the possibility to generate beneficial mutations that would be fixed, allowing them to become numerically dominant. The fact that only 4 of 12 populations evolved in this way could be related to the probability $\sim 1/3$ that the BNR events take place in the first 8h of daily evolution.

Let us notice that we are assuming very fast BNR ionization events, as compared with the bacterial motion. Only those bacteria placed along the ion shower are affected by it. We may estimate the duration of such a event from:

\begin{equation}
\tau_1\approx \frac{\epsilon\epsilon_0}{\sigma},
\end{equation} 

\noindent
where $\sigma=5.5\times 10^{-6} ~{\rm Coul/(V~s~m)}$ is the conductivity of pure water \cite{conduct}. That is, $\tau_1\approx 10^{-4}$ s. 

A second estimate for the duration comes from the difussion constants of ions in water \cite{D}, $D\approx 10^3~\mu m^2/s$. Taking $\lambda_D=0.5~\mu m$ as a characteristic dimension, results in:

\begin{equation}
\tau_2\approx \frac{\lambda_D^2}{D}\approx 2.5\times 10^{-4} s.
\end{equation} 

\noindent
In both cases, the times are of the order of $10^{-4}$ s. Taking into account that, at ambient temperatures, the typical speeds of bacterial motion are around 2 mm/s, only bacteria in contact with the ion shower, or very close to it, will be affected.

The fact that mutations with damage in the DNA repair mechanism are deleterious \cite{Lenski} is also consistent with the nature of BNR ionization processes. Indeed, the electron and ion shower is highly energetic and may produce such damages in the DNA, especially in the first steps after the n+p collision.

We shall compare the concentration of produced ions with the concentration of spontaneous ions in water, Eq. (\ref{n}). In each of the first 100 nm steps, the ejected proton creates around 300 ions. The induced concentration is, thus:

\begin{equation}
n_{ind}= \frac{300}{0.5^2\times 0.1\times {\rm \mu m^3}}=1.2\times 10^{16} {\rm ions/cm^3},
\end{equation} 

\noindent
that is, 4 times higher than $n$ given in Ec. (\ref{n}). The presence of ions in such high concentrations is also a strong mutagenic factor.

\section{Concluding Remarks}

In the present paper, we indicate a possible correlation between BNR ionization events and the LTEE observed rates of deleterious mutations with damages in the DNA repair and edit mechanism. In this way, we are indicating the probable origin of a class of ``spontaneous'' mutations.

The experimental confirmation of this possible correlation is plausible: restart the experiment by using fossils, and shield some of the evolving populations against the BNR. The shielded cultures should exhibit much lower rates for deleterious mutations with damages in the DNA repair mechanism. In around 1 - 2 years (2500 - 5000 generations), changes in mutation rates should be manifest.

On the other hand, a comment by Prof. Lenski \cite{Lenski} that some cancer cells also exhibit damages in the DNA repair mechanism, motivates us to rise the hypothesis about the BNR as one os the processes triggering cancer. Other events, like inhalation of radioactive Radon contained in air through breathing, are recognized carcinogens  \cite{Radon}. Defficient feeding, infectious proceeses, etc could be considered as conditions creating  an evolutive pressure over the expossed cells, similar to the limited amount of glucose in the LTEE. Under these conditions, the BNR induced deleterious mutations, with damages in the DNA repair and edit mechanism,  and the subsequent rise in the rate of spontaneous mutations, could allow the mutators to generate well adapted individuals that could become numerically dominant. In order to check this hypothesis, a controlled experiment in animals could be designed, for example in mouses, which are widely used as models of cancer in humans \cite{mice}.

\vspace{.5cm}
\begin{acknowledgements}
The author acknowledge C. Ceballos for the information on the role of BNR in microelectronics, and E. Altshuler, A. Cabo, C. Cruz, G. Mart\'in and E. Moreno for their comments and criticism. 
\end{acknowledgements}


\begin{thebibliography}{99}
\bibitem{IBM} J.F. Ziegler, H.W. Curtis, H.P. Muhlfeld et. al., IBM experiments in soft fails in computer electronics (1978-1994), IBM Journal of Research and Development, Vol. 40, No. 1, page 3, 1996. 
\bibitem{exp} T. Nakamura, M. Baba, E. Ibe, Y. Yahagi, H. Kameyama, Terrestrial
neutron-induced soft errors in advanced memory devices, 
Singapore, World Scientific, 2008; M. Nicolaidis, Ed., Soft errors in modern electronic systems, 
Springer, New York, 2010.
\bibitem{RFN} M.S. Gordon, P. Goldhagen, K.P. Rodbell, et. al., IEEE Transactions on Nuclear Science 51 (6), 3427 (2004). 
\bibitem{Lenski} R.E. Lenski, Phenotypic and genomic evolution during a 
20000 generation experiment with the bacterium E. Coli, in J. Janick, Ed., Plant Breeding
Reviews, Vol. 24, Part 2, page 225, 2004.
\bibitem{EELP} R. Lenski, Summary data from the long-term evolution experiment,
http://myxo.css.msu.edu/ecoli/summdata.html
\bibitem{pH} W. Boron, E.L. Boulpaep, Eds., Medical Physiology: A Cellular And Molecular Approach,
Elsevier, 2009.
\bibitem{sigma} J.W. Norbury, Nucleon-Nucleon Total Cross Section, NASA/TP-2008-215116.
\bibitem{sp} PSTAR : Stopping Power and Range Tables for Protons, 
http://physics.nist.gov/PhysRefData/Star/Text/PSTAR.html
\bibitem{Ip} NIST data on the ionization potential of water, http://webbook.nist.gov/cgi/cbook.cgi?ID=C7732185{\&}Mask=20
\bibitem{epsilon} M. Uematsu and E.U. Franck, J. Phys. Chem. Ref. Data, Vol. 9, No. 4, page 1291, 1980.
\bibitem{conduct} R.H. Shreiner and K.W. Pratt, Primary Standards and Standard Reference Materials for Electrolytic Conductivity, NIST Special Publication 260-142, 2004 Ed.
\bibitem{D} E.L. Cussler, Diffusion: Mass Transfer in Fluid Systems, New York, Cambridge University Press, 1997.
\bibitem{Radon} National Research Council. Committee on Health Risks of Exposure to Radon: BEIR VI. Health Effects of Exposure to Radon. Washington, National Academy Press, 1999.
\bibitem{mice} http://www.nih.gov/science/models/mouse/resources/hcc.html
\end{thebibliography}
\end{document}